\documentclass[iop]{emulateapj}
\usepackage{hyperref}
\usepackage{verbatim}

\begin{document}

\title{Modeling the Spectral Diversity of Quasars in the Sixteenth Data Release from the Sloan Digital Sky Survey}

\author{ Allyson Brodzeller\altaffilmark{1}}
\author{ Kyle Dawson\altaffilmark{1}}
\altaffiltext{1}{Department of Physics and Astronomy, University of Utah, Salt Lake City, UT 84112, USA}

\begin{abstract}

We present a new approach to capturing the broad diversity of emission line and continuum properties
in quasar spectra. We identify populations of spectrally similar quasars through pixel-level
clustering on 12,968 high signal-to-noise ratio (S/N) spectra from the Sloan Digital Sky Survey (SDSS) in the 
redshift range of $1.57<z<2.4$. Our clustering analysis finds 396 quasar spectra 
that are not assigned to any population, 15 misclassified spectra, and 6 quasars with incorrect 
redshifts. We compress the quasar populations into a library of 684 high S/N composite spectra,
anchored in redshift space by the Mg~\textsc{ii} emission line. Principal component analysis 
on the library results in an eigenspectrum basis spanning $1067 - 4007$~\AA. We model independent samples 
of SDSS quasar spectra with the eigenbasis, allowing for a free redshift parameter. Our models achieve 
a median reduced $\chi^2$ on non-broad absorption line quasar spectra that is reduced by 8.5\% relative to models using the
eigenspectra from the SDSS spectroscopic pipeline. A significant contribution to the relative 
improvement is from the ability to reconstruct the range of emission line variation. The redshift 
estimates from our model are consistent with the Mg~\textsc{ii} emission line redshift with an average
offset that displays 51.4\% less redshift-dependent variation relative to the SDSS eigenspectra. Our
method for developing quasar spectra models can improve automated classification and predict
the intrinsic spectrum in regions affected by intervening absorbers such as Ly$\alpha$, C~\textsc{iv}, 
and Mg~\textsc{ii}, thus benefiting studies of large-scale structure.
\newline
\end{abstract}

\section{Introduction}
Quasars exhibit distinct spectral features in the UV and optical wavelengths, featuring broad
and narrow emission lines with a continuum that can be approximated by a power law. The similarity permits 
co-addition of spectra to characterize general properties and amplify weaker spectral features
\citep[e.g.][]{francis91,vandenberk01,tammour15,harris16,jensen16}. While the first-order 
similarity is impressive, emission line profiles and continuum slopes vary across the full
population and in time. Reverberation mapping provides insight into the structure of the emitting 
regions and the central black hole \citep[see][for example]{shen15,kaspi21,yu21}; however, a 
comprehensive model that explains the full range of diversity is lacking 
\citep[e.g.][]{ruan14,luo15,dyer19,matthews20}. 

Complications in spectral modeling stem from variation at all wavelengths. Within the same quasar 
spectrum, the locations of different emissions lines are offset nonuniformly from the systemic
redshift \citep[e.g.][]{hewett10,shen16}. Across different spectra, emission lines exhibit 
varying degrees of asymmetries in their profiles. Some quasars also display broad, blue-shifted 
absorption troughs associated with prominent emission lines, referred to as broad absorption line
(BAL) quasars \cite[e.g.][]{weymann91,hall02}. The number of BALs in a spectrum, along with their
corresponding velocity widths and blueshifts, varies from one BAL quasar to the next.

Without an accepted theoretical model, modeling efforts rely on empirical trends in spectral
properties and correlations between features. For example, the equivalent width of broad emission
lines is known to be anticorrelated with continuum luminosity, referred to as the Baldwin Effect. 
The anticorrelation was first observed with the C~\textsc{iv} emission line \citep{baldwin77} and 
later confirmed in other lines \citep[see][for a review]{shields07}. The physical mechanism that
drives the Baldwin Effect is still debated. Principal Component Analysis (PCA) is a mathematical
tool that reveals significant sources of signal variance and empirical correlations between
features. \citet{boroson92} performed PCA on measured properties of low redshift quasars, revealing 
a strong relationship between the strengths of Fe~\textsc{ii} and [O~\textsc{iii]} emission and 
properties of the H$\beta$ emission line, known as Eigenvector 1 (EV1). EV1 trends are believed to
be related to the Eddington Ratio \citep[e.g.][]{boroson02,shen14}.

PCA implemented directly on quasar spectra captures intrinsic variance on the level of individual flux
densities. Eigenspectra covering the UV wavelengths are connected to the Baldwin Effect 
\citep[e.g.][]{shang03,yip04} and reveal that significant variance can be attributed to the cores of broad 
emission lines and continuum slopes \citep[e.g.][]{francis92,rochais17}. In the first application of
PCA on quasar spectra, \citet{francis92} demonstrated that linear combinations of a small number of
eigenspectra can adequately reconstruct observed data from the Large, Bright Quasar Survey 
\citep{hewett95}. PCA has since become a popular method for developing data-driven models of 
quasar spectra \citep[e.g.][]{suzuki05,suzuki06,paris11,paris12,davies18,guo19}. A single set of four
eigenspectra was used for automatic identification and redshift estimation for all quasar spectra 
in the Baryon Oscillation Spectroscopic Survey \citep[BOSS;][]{dawson13} and the extended Baryon Oscillation
Spectroscopic Survey \citep[eBOSS;][]{dawson16}, as described in \citet[][hereafter B12]{bolton12}.
\citet{yip04}, however, argue that because spectral diversity is dependent on both redshift and luminosity
\citep[e.g.][]{vandenberk04,jensen16}, there does not exist a compact set of quasar eigenspectra
that adequately describes all quasars. The authors suggest that eigenspectra developed for specific ranges in
this parameter space are necessary for accurate modeling.

In this work, we implement a technique that clusters quasar spectra in the redshift range $1.57<z<2.4$
based on similarity at the pixel level. We use quasar spectra from the Sixteenth Data Release
\citep[DR16;][]{ahumada20} of the fourth generation of the Sloan Digital Sky Survey 
\citep[SDSS-IV;][]{blanton17}. Potential contaminants and peculiar quasars in the sample are easily 
discovered, as highly dissimilar spectra naturally segregate. We use the clusters of similar quasar 
spectra to construct a library of high signal-to-noise ratio (S/N) composite spectra that is representative of 
the spectral diversity across the entire sample. We perform PCA on the library and compare the 
modeling performance of the resulting eigenspectra to that of the B12 eigenspectra. Our method for 
developing quasar eigenspectra differs from previous work, as PCA is performed not on a selection 
of individual quasar spectra but rather on composite spectra built from quasar populations with low
internal diversity, promoting representation of recurrent spectral features. In addition, the high S/N of the composite spectra allows us to apply a redshift correction to the PCA training sample through measurement of the Mg~\textsc{ii} emission line location and therefore mitigate potential offsets.

In Section~2, we describe the training sample of quasar spectra selected for our clustering analysis.
We also detail the quasar samples used for testing the modeling performance of the eigenspectra. In 
Section~3, we introduce our clustering technique, present the clustering results, and outline how we transform 
clusters of similar spectra into a library of high S/N composite spectra. In Section~4, we
perform PCA on the library and present our eigenspectra. We select the number of eigenspectra for 
modeling spectra of the test samples and compare the modeling performance to B12 eigenspectra. 
In Section~5, we summarize our results from this work and outline future improvements with data from 
the Dark Energy Spectroscopic Instrument \citep[DESI;][]{DESIcollaboration16a,DESIcollaboration16b}.

\section{Data}
This work uses quasar spectra and their classifications from DR16 of SDSS-IV collected
as a part of the BOSS and eBOSS programs. A primary goal of BOSS and eBOSS was to constrain
cosmological parameters through measurements of the baryon acoustic oscillation feature in
the distribution of matter. The final cosmological implications from BOSS and eBOSS are 
summarized in \citet{eBOSSfinal}. 

Quasars were a primary spectroscopic target of BOSS and eBOSS. At $z<2.2$, quasars 
are used as direct tracers of the matter distribution \citep{neveux20,hou21}, while 
at higher redshifts the Ly$\alpha$ forest observed as absorption in quasar spectra 
traces the matter distribution \citep{bourboux20}. We select quasar samples for this 
study from the final quasar catalog of DR16 \citep[DR16Q;][]{lyke20}. This section 
summarizes the spectroscopic data and selection criteria for the training sample
used in our clustering analysis. We then describe the calibration and filtering 
procedures applied to the training sample to maximize the information gained from 
clustering. Finally, the test samples for evaluating the performance of our eigenspectra
models are presented.

\subsection{Spectroscopic Data}
\label{subsec:SpecData}

The BOSS and eBOSS programs used a pair of identical multi-object fiber spectrographs 
\citep{smee13} on the 2.5 m Sloan telescope \citep{gunn06} at Apache Point Observatory. 
Observations were taken using aluminum plates, each of which subtended $3\degr$ on the
sky. Each plate featured 1000 holes drilled $2\arcsec$ in diameter at locations corresponding 
to calibration, stellar, galaxy, or quasar targets. Once an observation with a plate was 
completed, the 1000 spectra were processed and classified \citep{bolton12}. 
The method for selecting quasar targets in BOSS and eBOSS is detailed in \citet{ross12} 
and \citet{myers15}, respectively.

\subsection{Training Sample}
\label{subsec:TrainingSample} 

We apply a clustering technique to a training sample of quasar spectra to identify 
groups that possess similar spectral features. The training sample must capture 
the diversity of continuum and emission line properties to adequately represent the
entire quasar population. Additionally, the training sample must be relatively 
free of nonintrinsic spectral features to increase the true physical similarity
within clusters. These two factors are the primary considerations when selecting the 
training sample. 

The training sample is selected based on redshift, balnicity, and S/N of 
the spectra. We use the \verb|Z_PCA| redshifts in DR16Q, which were estimated
using the four B12 eigenspectra \citep[see Section~4 of][]{bolton12} plus a three-term polynomial in $\log\lambda$ for broadband. Only quasars 
with reliable redshift estimates are included in the training sample by requiring
\verb|ZWARN_PCA = 0|. Quasar spectra containing BALs are excluded by requiring a 
\verb|BAL_PROB| attribute of zero. This exclusion ensures that clusters form based on 
intrinsic similarity. Finally, the average S/N ratio per pixel is determined
over the entire observed spectrum after excluding pixels identified as having poor sky
subtraction. A minimum average S/N ratio of 10 per pixel is required for the
training sample.

The redshift range is selected to maximize the number of spectral features covered by the
training sample. The identification of BALs requires the presence of the C~\textsc{iv} emission 
line at the rest-frame wavelength of 1549~\AA, leading to a minimum redshift of $z=1.57$. 
The upper limit in redshift is set to $z=2.4$ by requiring that the Mg~\textsc{ii} emission line be present 
at a reasonably high S/N. The resulting training sample contains 12,968 quasars.
As shown in Table~\ref{table_bins}, the training sample is binned into four discrete subsets 
of width $\Delta z \approx 0.2$. Clustering is performed within each redshift bin independently
to account for potential redshift evolution.

All spectra within a redshift bin are cropped to a common rest-frame wavelength range
defined from $\lambda_{min}$ to $\lambda_{max}$. $\lambda_{min}$ is set by the bluest
pixel of the lowest redshift spectrum in the bin, and $\lambda_{max}$ is set by the
reddest pixel in the highest redshift spectrum in the bin. The flux of each spectrum is
normalized according to its median computed over the common wavelength range. The rest-frame
values of $\lambda_{min}$ and $\lambda_{max}$ for the redshift bins are reported in
Table~\ref{table_bins}.

\begin{table}[htb]
\centering
\caption{The number of quasars in each redshift bin and wavelength range used for clustering}
\begin{tabular}{c | c c c} 
 \hline
 Redshift Range & Quasars & $\lambda_{min}$ (\AA) & $\lambda_{max}$ (\AA) \\ 
 \hline\hline
$1.57<z<1.8$ & 3201 & 1396.69 & 3714.50 \\ 
$1.8<z<2.0$ & 3393 & 1273.21 & 3460.19 \\
$2.0<z<2.2$ & 3006 & 1189.60 & 3233.70 \\
$2.2<z<2.4$ & 3368 & 1115.32 & 3057.03 \\ 
 \hline
\end{tabular}
\label{table_bins}
\end{table}

As a simple diagnostic of diversity, we explore the distribution of equivalent width of
the C~\textsc{iv} emission line within each redshift bin. The equivalent width is determined 
through fitting a double Gaussian to the continuum-subtracted flux in the range $1510 - 
1590$~\AA. The continuum flux in the region of the C~\textsc{iv} emission line is estimated by 
fitting a power law over the ranges $1430 - 1450$~\AA\ and $1680 - 1700$~\AA. The location, width,
and amplitude parameters for each Gaussian vary independently as expected from kinematics of 
gas responsible for narrow and broad emission. The equivalent width is calculated only for 
C~\textsc{iv} emission regions containing 20 or more reliable pixels. The mean and standard 
deviation of the equivalent width for the bins are $24.87 \pm 11.76$, $24.23 \pm 10.84$, 
$23.41 \pm 10.62$, and $22.61 \pm 10.21$, in order of increasing redshift. Due to selection
effects, higher luminosity objects are more likely to be observed with increasing redshift; thus, 
the apparent decrease in equivalent width is the Baldwin Effect for C~\textsc{iv}. 

\begin{figure}[htb]
  \centering
	  \includegraphics[width=0.45\textwidth, angle=0]{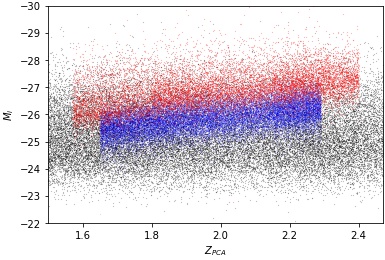}
	  \caption{
	  The black data points correspond to the absolute i-band magnitude
	  of a 10\% random sampling of all quasars in DR16Q as a function of redshift.
	  Only quasars detected at $1\sigma$ in the i-band are displayed. The quasars
	  selected for the training sample are shown in red. The quasars selected for
	  the standard test sample are shown in blue. The vertical lines
	  indicate the redshift bins of the training sample.}
	  \label{fig:Mi_z}
\end{figure}

The distribution of absolute magnitude in the i-band as a function of redshift for quasars
in DR16Q is shown in Figure~\ref{fig:Mi_z}. The training sample is highlighted in red. 
The training selection is intrinsically more luminous than the general population owing to 
the S/N requirement. In addition, the training sample trends toward higher 
luminosity with increasing redshift. This trend explains the decreasing equivalent width 
of the C~\textsc{iv} emission line with increasing redshift that was described in the 
previous paragraph.

The quasar spectra in the training sample are calibrated to minimize spurious signal.
All spectra are corrected for galactic extinction using the Fitzpatrick model 
\citep{fitzpatrick99} and dust extinction map from \citet*{schlegel98}. Pixels flagged
by the eBOSS pipeline as contaminated by sky emission are masked. Flux measurements at 
$\lambda_{obs} < 3610.5$~\AA\ are removed owing to uncertainty in flux and wavelength 
calibration. Each spectrum is then shifted to its rest-frame wavelength solution based on
the \verb|Z_PCA| redshift estimate. Pixels in the Ly$\alpha$ forest at 
$\lambda_{em} < 1216$~\AA\ are masked to eliminate stochasticity arising from
external sources.

\begin{figure}
  \centering
	  \includegraphics[width=0.45\textwidth, angle=0]{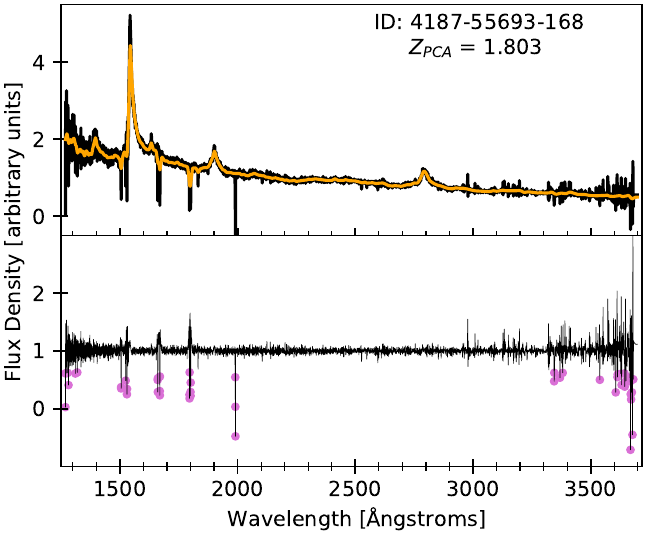}
	  \caption{An example of the NAL filter
	  applied to a quasar spectrum.
	  \textbf{Top:} the measured flux density is shown in black, with the
	  estimated spectrum from the Gaussian smoothing filter overlaid
	  in orange. The ID code corresponds to the unique PLATE-MJD-FIBERID
	  of every SDSS spectrum. \textbf{Bottom:} the ratio of measured
	  flux to smoothed flux is shown in black. 
	  Pixels at which the ratio deviates negatively by greater than three standard deviations 
	  are marked in purple. These flux decrements likely originate from CGM 
	  absorption and are masked in our clustering analysis.}
	  \label{fig:absLines}
\end{figure}

We next identify and remove narrow absorption lines (NALs) in the quasar spectra. NALs are
imprinted by diffuse circumgalactic medium (CGM) along the line of sight \citep[see][]{tumlinson17}.
The intrinsic spectrum of each quasar is estimated by applying an error-weighted 
Gaussian smoothing filter with a dispersion of 8 pixels. NALs are identified at 
pixels where the ratio of measured flux to smoothed flux deviates negatively by more 
than three standard deviations. Pixels with suspected absorption lines are masked. An 
application of the NAL filter is shown in Figure~\ref{fig:absLines}.

\subsection{Standard Test Sample}
\label{subsec:TestSample}

A clustering analysis applied to the training sample is used to derive quasar eigenspectra
(discussed in Section~\ref{subsec:eigenspectra}). We evaluate the ability of these eigenspectra 
to model an independent test sample of quasar spectra. The standard test sample is selected 
based on redshift and S/N requirements, excluding quasars with a nonzero
\verb|BAL_PROB| attribute. The redshift range of the standard test sample is cropped to 
$1.65<z<2.29$ to allow for a free redshift parameter in the fitting. Only spectra with a 
\verb|ZWARN_PCA| attribute of zero are included. Flux measurements at rest-frame wavelengths 
shorter than 1216~\AA\ according to \verb|Z_PCA| are given a weight of zero to mask the
Ly$\alpha$ forest. The quasar spectra in the standard test sample have an average S/N
ratio per pixel over the entire spectrum ranging between 5 and 10. The upper bound in 
S/N is imposed to avoid overlap with the training sample. 

The standard test sample contains 26,498 quasars. The distribution of absolute magnitude
in the i-band for the standard test sample as a function of redshift is highlighted in
blue in Figure~\ref{fig:Mi_z}. As with the training sample, the S/N requirement 
leads to intrinsically brighter quasar sources than the general population and shows
a trend of increasing luminosity with redshift.

\subsection{BAL Test Sample}
\label{subsec::BalTestSample}

BALs are observed in a significant fraction of quasar spectra. In the UV and visible
wavelengths, the fraction is reported between 10\% and 30\% 
\citep[e.g.][]{trump06,paris12,guo19}. Recent work \citep[e.g.][]{rankine20} supports 
the theory that quasars containing BALs are not a distinct class of quasars. Rather,
the BALs most likely arise as a result of orientation effects that place outflowing 
gas along the line of sight, with all quasars having some probability to observed with BALs.
We test the ability of our eigenspectra to model the unabsorbed spectra of a sample
of BAL quasars.

The BAL test sample is selected from the \citet{guo19} BAL Quasar Catalog. The quasars in 
the catalog are from DR14 of SDSS-IV \citep{abolfathi18} and were classified as BAL quasars 
using a convolutional neural network. The catalog contains the \verb|BAL_PROB| determined by 
the neural network and the most robust redshift estimate for each quasar, \verb|Z|. 
Only objects with a \verb|BAL_PROB| of greater than 0.5 are included in the BAL Quasar Catalog.
Quasars within the redshift range $1.65<z<2.29$ are selected for our BAL test sample. The BAL
features recorded in the catalog are measured relative to \verb|Z| so the sample is selected 
using this redshift estimate rather than \verb|Z_PCA|. We set a minimum average S/N ratio 
per pixel requirement of 5. We do not impose a maximum value on the S/N ratio per
pixel, as all quasars in the training sample have a \verb|BAL_PROB| of zero, and thus there is no 
overlap between the two samples. The final BAL test sample contains 12,431 quasars.

The BAL Quasar Catalog includes both the blueshifts and velocity widths of absorption troughs 
associated with the C~\textsc{iv} and Si~\textsc{iv} emission lines. This information is used
to mask the regions of a spectrum contaminated by BALs with velocity widths greater than 450
km s$^{-1}$. In addition to masking the BAL troughs, the Ly$\alpha$ forest at 
$\lambda_{em} < 1216$~\AA\ is masked. We then model the remaining unmasked pixels with our 
eigenspectra. 

\section{Clustering}
We use a clustering technique as a form a data compression. We transform 12,968 individual spectra
into a library of composite spectra built from clusters of quasars with similar spectra. In this 
section, we discuss our clustering method and the results within each redshift bin. We then describe
the objects that fail to cluster and examine possible causes. Lastly, we outline the procedure for 
transforming the clusters into a library of high S/N composite quasar spectra corrected
to the rest-frame wavelength of the Mg~\textsc{ii} emission line.

\subsection{SetCoverPy}
\label{subsec:setcoverpy}

SetCoverPy\footnote{\url{https://github.com/guangtunbenzhu/SetCoverPy}} \citep{zhu16} is a 
classification algorithm based on the Set Cover Problem (SCP). A full description of SetCoverPy and 
the SCP is beyond the scope of this paper, and we refer the reader to \citet{zhu16} for more details. 
In brief, SetCoverPy identifies representatives of a sample based on a similarity (or distance) metric
between instances. Instances with strong similarity (or a small distance) are able to represent each 
other. The threshold for two instances to be considered similar enough such that they can represent 
each other is a free parameter determined by the user. The goal of SetCoverPy is to find an optimal set
of instances that can represent the entirety of the sample at the minimum cost. These representative
instances are dubbed archetypes. If all instances have uniform cost to be labeled an archetype, the 
solution is the minimum number of archetypes necessary to represent the sample. A single instance may 
be represented by multiple archetypes or be an archetype only for itself. 

Implementing SetCoverPy requires selecting a distance metric to measure the similarity of instances and
defining the cost for each instance to be labeled an archetype. Given a minimum distance within which 
pairs of instances can be considered similar and a pairwise distance matrix, SetCoverPy returns a list
of archetypes of the input sample. We use SetCoverPy with a few modifications on our training sample of
quasars to identify clusters with similar spectra.

\subsection{SetCoverPy on Quasar Spectra}
\label{subsec:SetCoverQSO}

We find clusters of similar quasar spectra in each redshift bin of the training sample using spectral
archetypes identified with SetCoverPy. We measure the similarity of quasar spectra using the reduced
$\chi^2$ statistic (see Section~\ref{subsec:distance}). All quasar spectra are assigned equal
cost; thus, SetCoverPy returns the minimum number of archetypes necessary to represent every spectrum
in a redshift bin. We treat quasar archetypes as cluster centers around which we build clusters. The 
cluster members are assigned based on the minimum distance used to determine the archetypes.

\begin{figure*}[htb]
    \centering
        \includegraphics[width=0.85\textwidth, angle=0]{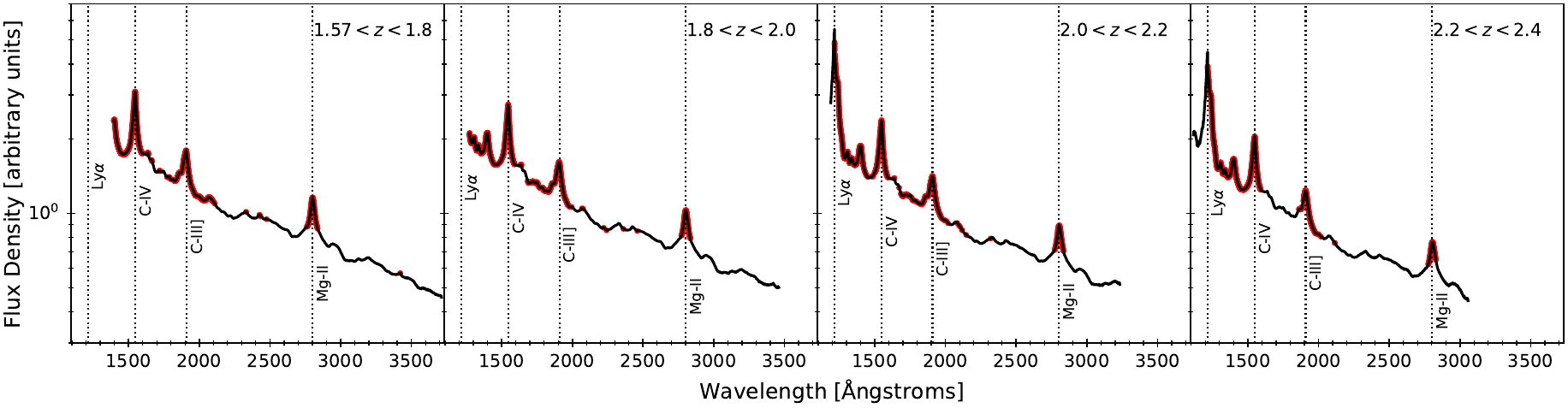}
        \caption{The error-weighted mean spectrum of each 
        redshift bin is shown in black. The pixels that exhibit spectral
        diversity in the 68th percentile or higher are shown in red. Pixels at
        wavelengths $<1216$~\AA\ are excluded from the percentile cut
        due to the presence of the Ly$\alpha$ forest.
        The rest-frame wavelengths of Ly$\alpha$, C~\textsc{iv}, C~\textsc{iii]}, and Mg~\textsc{ii} emission are indicated by
	    vertical lines.}
        \label{fig:68perc}
\end{figure*}

A single quasar spectrum may fall within the minimum distance to multiple archetypes. The quasar is
assigned to multiple clusters in this case. We find this overlap feature especially desirable because
there are no clear boundaries from one subclass of quasars to the next. Rather, there exist natural
gradients and/or overlap between populations. Deciding where to draw the boundary line along these 
gradients is frequently arbitrary and may lead to the misclassification of near-boundary objects 
\citep[for an example of this problem with galaxy populations, see][]{pandey20}.

Further, there are individual quasar spectra that are archetypes only for themselves, not falling
within the minimum distance to any other spectrum. We define quasars that are archetypes only for 
themselves as cluster outliers. We introduce an iterative procedure for reducing the number of cluster 
outliers and consequently increasing diversity coverage in the library of composite quasar spectra. 
This procedure is detailed in Section~\ref{subsec:IterativeClustering}.

\subsubsection{Distance Metric}
\label{subsec:distance}

We adopt the reduced $\chi^{2}$ statistic as the distance metric for measuring spectral similarity
in the training sample.
The $\chi^2$ statistic is analogous to a weighted squared Euclidean distance in N-dimensional space.
The reduced $\chi^{2}$ between two quasar spectra in a redshift bin is computed as   
\begin{equation}
\label{eq:chi2red}
    \chi_{ij}^2 = \frac{1}{N-2}\sum_{n=1}^{N}\frac{(f_i(\lambda_n) - c\lambda^{\Delta\alpha_\lambda}f_j(\lambda_n))^2}{\sigma_i(\lambda_n)^2 + (c\lambda^{\Delta\alpha_\lambda}\sigma_j(\lambda_n))^2}
\end{equation}
where N is the number of pixels covered by the common wavelength range that have reliable measurements 
for both spectra ($\frac{1}{\sigma_i^2\sigma_j^2} > 0$). In pixel-level clustering, the difference in
power-law spectral indices between two spectra in a pair will be present across all pixels and can
dominate the $\chi^2$. The $c\lambda^{\Delta\alpha_\lambda}$ term accounts for this difference, 
effectively removing the power-law component of the continuum from consideration in the distance metric. 
The two parameters, c and $\Delta\alpha_\lambda$, are determined by minimizing the reduced $\chi^2$ in 
every pairwise comparison of quasar spectra. Symmetry is maintained in the distance matrix by always 
applying this term to the higher S/N spectrum of the pair.

We found that computing the reduced $\chi^2$ using every pixel in the common wavelength range allows
continuum-dominated portions of the spectra to dilute the distance statistic.
We therefore focus the reduced $\chi^2$ statistic on the pixels that exhibit the most flux 
diversity in each redshift bin to increase the information content. These pixels were identified 
by computing the average reduced $\chi^2$ about the error-weighted mean using all spectra in the 
bin. Those pixels in the 68th percentile and above are highlighted in the mean spectrum of each bin
in Figure~\ref{fig:68perc}. Only the highlighted pixels factor into the reduced $\chi^2$ statistic
between quasar spectra pairs. This selection of pixels focuses the comparison primarily on the 
prominent emission features. The pixels blueward of $1216$~\AA\ do not factor into determining the
68th percentile cut on diversity, as these are masked owing to the presence of Ly$\alpha$ forest.

\subsubsection{Iterative Clustering}
\label{subsec:IterativeClustering} 

\begin{table*}[htb]
\centering
\caption{Final Clustering Statistics}
\begin{tabular}{c| c c c c c c} 
\hline
Redshift & Initial & Number of & Largest & Average & Initial Number & Final Number  \\ 
Bin & Reduced $\chi^2$ & Clusters & Cluster Size & Cluster Size & of Outliers & of Outliers\\
 \hline\hline
$1.57<z<1.8$ & 1.071 & 194 & 822 & 110 & 547 & 34   \\
$1.8<z<2.0$ & 1.147 & 183 & 1169 & 164 & 625 & 71   \\ 
$2.0<z<2.2$ & 1.350  & 152 & 1001 & 166 & 639 & 92  \\
$2.2<z<2.4$ & 1.573 & 160 & 1191 & 182 & 914 & 206  \\
 \hline
\end{tabular}
\label{table_stats}
\end{table*}

\begin{figure*}[htb]
    \centering
        \includegraphics[width=0.85\textwidth, angle=0]{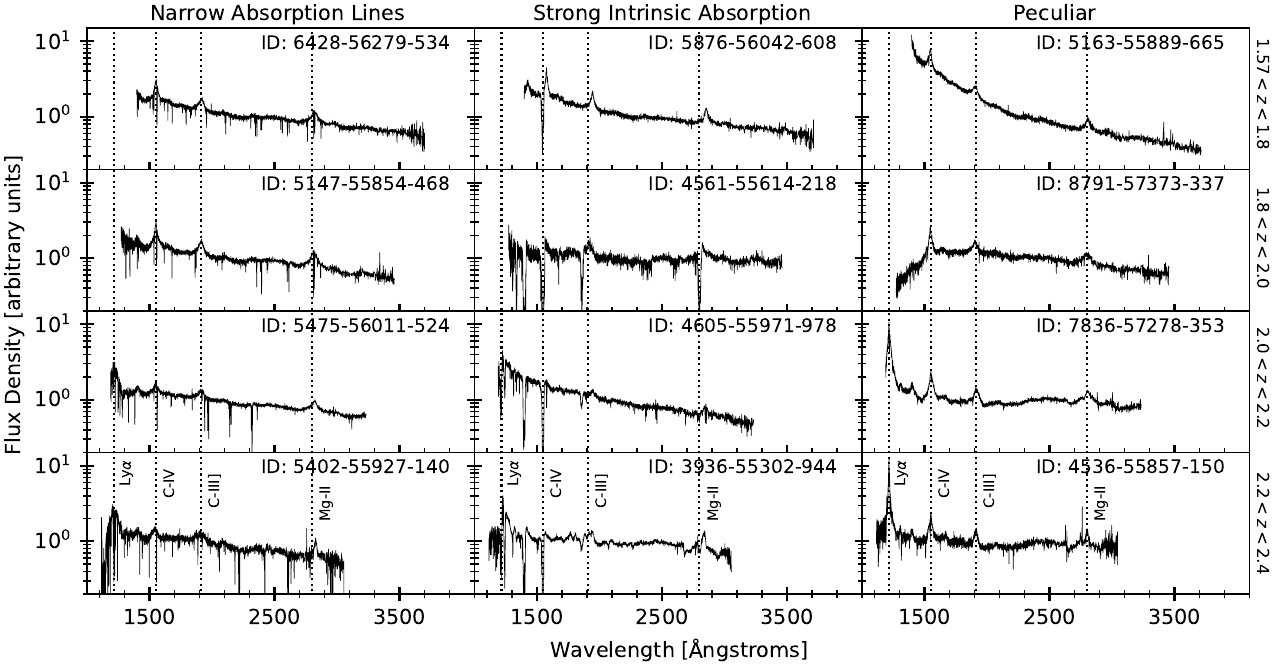}
        \caption{
        Example spectra from each redshift bin that belong to one of the three primary outlier classes: spectra with unmasked NALs imprinted by CGM,
        spectra with strong intrinsic 
        absorption, and spectra with peculiar shapes or features such as the flux calibration error seen in the peculiar spectrum of the highest redshift bin.
        The spectra have been shifted to the rest-frame wavelength
        solution according to the Z\_PCA redshift estimate. 
        Quasar spectra with strong intrinsic absorption
        often also have an incorrect redshift
        estimate.}
\label{fig:outlierExamples}
\end{figure*}

The minimum distance within which quasar pairs can be considered similar is the only free 
parameter in SetCoverPy; therefore, its selection dictates the solution. A larger minimum 
distance will reduce the number of archetypes necessary to represent the entire sample of
quasars. Conversely, a smaller minimum distance results in a larger number of clusters with
less internal diversity. In addition, as the minimum reduced $\chi^2$ is decreased, more 
quasars will be identified as cluster outliers. 

We select the minimum reduced $\chi^2$ for each redshift bin by balancing the number of cluster 
outliers and the size of the largest cluster. To avoid overfitting the minimum distance parameter,
each redshift bin is randomly subsampled three times. SetCoverPy is implemented on each random 
sample over a range of minimum reduced $\chi^2$ values. The fraction of quasars labeled as outliers
and the fraction of quasars assigned to the largest cluster are recorded in each iteration. We choose
the minimum reduced $\chi^2$ value for each redshift bin where the averages of these two
fractions over the three subsamples are approximately equal. The equality occurs at the reduced $\chi^2$
values given in Table~\ref{table_stats}. The minimum reduced $\chi^2$ increases with redshift.

The resulting outlier fraction in each redshift bin lies between 15\% and 30\%. Neglecting all 
of the outliers reduces the diversity coverage of clusters. Inspection of the cluster outliers 
reveals that many fall on the boundary of the minimum distance cutoff from one or more of the 
archetypes. We preserve the initial cluster structure and retrieve outliers by adding them to 
existing clusters in an iterative process. The process also permits outliers to form new clusters.

The outlier fraction is reduced by incrementally increasing the minimum reduced $\chi^2$ by 
10\% of the original value for the redshift bin. SetCoverPy is run on a data set containing
only the archetypes and cluster outliers with the increased minimum distance value. The quasar 
archetypes are assigned a cost of zero since they have already been assigned as cluster centers.
The outliers are assigned a uniform cost of one. Any outlier that can be
represented by an existing archetype(s) in the new run of SetCoverPy is added to the cluster(s) 
centered on that archetype(s). If an outlier is determined to be an archetype of the updated data 
set, representing N other outliers, it is added to the list of archetypes. The outliers within 
the minimum distance to the new archetype then form a cluster. We repeat this procedure by
increasing the minimum reduced $\chi^2$ another 10\% from the original value and implementing 
SetCoverPy on the updated list of archetypes and remaining outliers. The minimum reduced $\chi^2$
value is increased until it reaches 150\% the original value. The results of clustering in each bin
are detailed in Table~\ref{table_stats}.

\subsubsection{Outliers}
\label{subsec:outliers}

We identify three broad categories of outliers through visual inspection of their spectra. 
Outlier spectra most commonly have several unmasked NALs, strong intrinsic absorption features
from high column density associated absorbers, or truly peculiar spectral shapes or features. 
Outlier spectra also result from errors in flux calibration.
In Figure~\ref{fig:outlierExamples}, we show examples of outlier quasar spectra that fall 
into each of the three primary categories. The PCA redshift estimate is often incorrect for quasars 
in the first two of these categories. The incorrect redshift estimates are particularly 
evident in a few examples of the outlier quasar spectra with strong intrinsic absorption.
We also find five outlier spectra from the highest redshift bin with incorrect redshifts where the
C~\textsc{iv} emission line was mistaken for Ly$\alpha$ in the PCA fit and one outlier in the
lowest redshift bin where the C~\textsc{iii]} emission line was mistaken for C~\textsc{iv}. 
Lastly, we find one outlier spectrum that is a misclassified galaxy.

The outlier spectra that exhibit NALs from CGM indicate a shortcoming of the NAL filter
(see Section~\ref{subsec:TestSample}). The
filter fails to adequately estimate the intrinsic spectrum in these cases. Visual inspection
reveals that the vast majority of the quasar spectra with unmasked CGM absorption are not 
intrinsically peculiar. Rather, the spurious absorption features drive the pairwise reduced
$\chi^2$ to higher values. Further, the occurrence of outliers with spectra featuring heavy 
CGM absorption increases with redshift. From the low to high redshift bin, we identify 2, 8,
32, and 80 cluster outliers as having an excess of unmasked NALs. In future 
studies, a more aggressive approach to identifying and removing NALs may be necessary.

\begin{figure}[htb]
  \centering
	  \includegraphics[width=0.45\textwidth, angle=0]{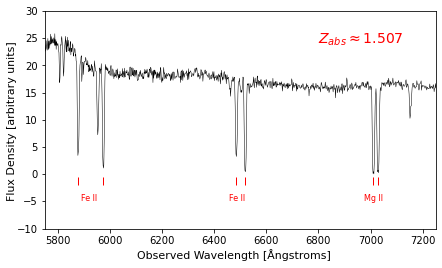}
	  \includegraphics[width=0.45\textwidth, angle=0]{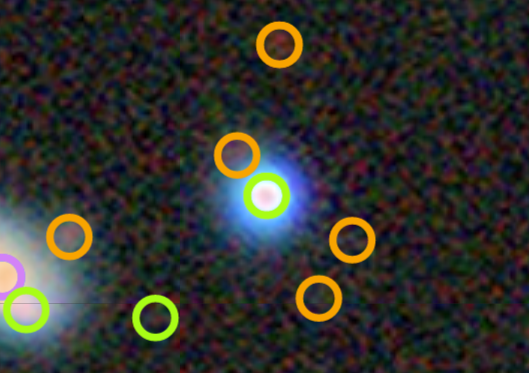}
	  \caption{\textbf{Top:} a region of the observer-frame spectrum of an outlier quasar at $z=2.028$
	  featuring a candidate absorption system at $z\approx1.507$.
	  \textbf{Bottom:} imaging data from DR9 of the DESI Legacy Imaging
	  Survey centered on the quasar. Sources 
	  in DR9 are indicated with circles. There is a source to the northwest
	  of the quasar that may be associated with the absorption
	  seen in the spectrum.}
	  \label{fig:legacySurvey}
\end{figure}

An unintentional benefit of our clustering method is finding quasar spectra that feature 
absorption from several intervening systems with ease, as they naturally segregate from the rest of
the population. From a sample of 12,968, roughly 120 quasars are 
automatically identified as having extreme CGM absorption. Quasar spectra featuring this
type of absorption are important tools for studying galaxies and their local environments 
\cite[][for example]{tumlinson17,lundgren21,zou21}. Following the work of \citet{zhu13}, we
identify a discrete absorption system in the quasar spectrum shown in the first column, third row 
of Figure~\ref{fig:outlierExamples} by the presence of doublet Mg~\textsc{ii} and four strong
Fe~\textsc{ii} (2344, 2383, 2586, 2600~\AA) absorption lines coincident at a redshift
of $z\approx1.507$. We show the observer-frame region of this spectrum that features the absorption
system in Figure~\ref{fig:legacySurvey} along with imaging data from DR9 of the DESI
Legacy Imaging Survey\footnote{https://www.legacysurvey.org/} \citep{dey19}. The circles in 
the imaging data indicate sources in DR9, colored by morphological type. The orange circles 
are round exponential galaxies, the green circles are point sources, and the purple circles are sources with
Sersic profiles. The image reveals a potential galaxy associated with absorption located 2\farcs48 to the northwest of the quasar.

In addition to the typical outliers, five clusters incorporating 14 total spectra were identified 
as abnormal. This includes two clusters in redshift bin $1.57<z<1.8$, two clusters in redshift bin
$2.0<z<2.2$, and one cluster in redshift bin $2.2<z<2.4$. The spectra in these clusters 
display a power-law continuum but contain no emission features. It is likely that these clusters 
originated from misclassified stellar spectra, as there are no obvious Mg~\textsc{ii} absorption 
features. These clusters are rejected from the final library of composite quasar spectra.

\subsection{Composite Quasar Spectra}
\label{subsec:composites} 

An error-weighted mean composite spectrum is constructed for all clusters using observer-frame
wavelengths longer than 3610.5~\AA. Prior to averaging, pixels contaminated by sky emission are
masked, all spectra are shifted to the rest-frame solution given by \verb|Z_PCA|, and the spectra
are corrected for galactic extinction and the effective optical depth of the Ly$\alpha$ forest using
the model from \citet{kamble20}. All spectra are normalized over a wavelength range of $1500 - 
2950$~\AA, which is common to all redshift bins. This normalization ensures physicality in modeling 
the continuum across all redshifts. The NAL filter introduced in Section~\ref{subsec:TestSample} 
is applied to the individual spectra to reduce the influence of CGM absorption. Finally, the two-parameter 
warping term, $c\lambda^{\Delta\alpha_\lambda}$, is applied to match the spectral index of each cluster 
member to that of the archetype spectrum. The composites are built including flux measurements beyond
the common rest-frame wavelength range of the redshift bin. The inclusion of flux measurements beyond
this range assumes that quasars with similar spectra within the common wavelength range likely have 
similar spectra beyond.

Redshift corrections are applied to all the cluster composite spectra using the Mg~\textsc{ii} emission line
to correct for potential offsets in the \verb|Z_PCA| redshift estimates. The Mg~\textsc{ii} emission line is
selected as the reference for its low bias relative to the systemic redshift \citep{shen16}. The 
continuum flux in the region of the Mg~\textsc{ii} emission line is estimated for each composite spectrum by 
fitting a power law to the wavelength ranges of $2750 - 2765$~\AA\ and $2840 - 2855$~\AA. A Gaussian is 
fit to the continuum-subtracted flux in the range of $2770 - 2835$~\AA. The spectrum is then shifted 
such that the Gaussian peak aligns with 2799.5~\AA, the effective center of the Mg~\textsc{ii} doublet. The 
mean redshift correction is 412 km s$^{-1}$ with a standard deviation of 607 km s$^{-1}$.

\begin{figure}[htb]
  \centering
	  \includegraphics[width=0.45\textwidth, angle=0]{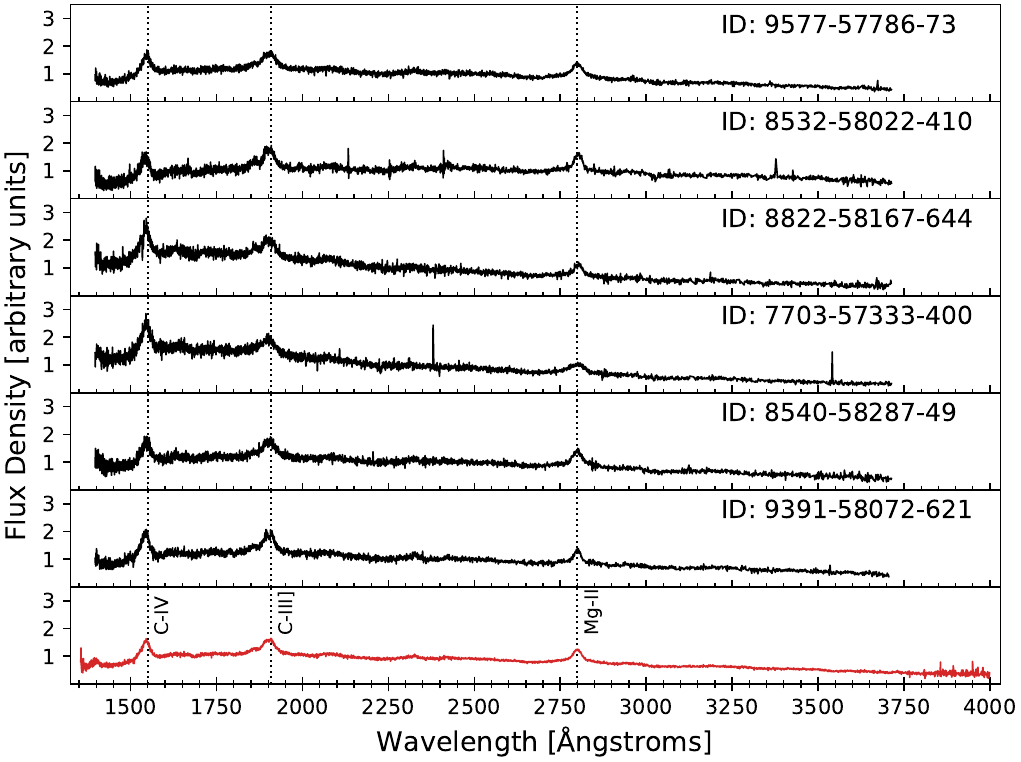}
	  \caption{The six quasar spectra shown in black clustered together in 
	  redshift bin $1.57<z<1.8$. The first panel shows the archetype spectrum.
	  The resulting composite spectrum for the cluster
	  is shown in red in the bottom panel. }
	  \label{fig:buildEx}
\end{figure}

\begin{figure*}[htb]
    \centering
        \includegraphics[width=0.93\textwidth, angle=0]{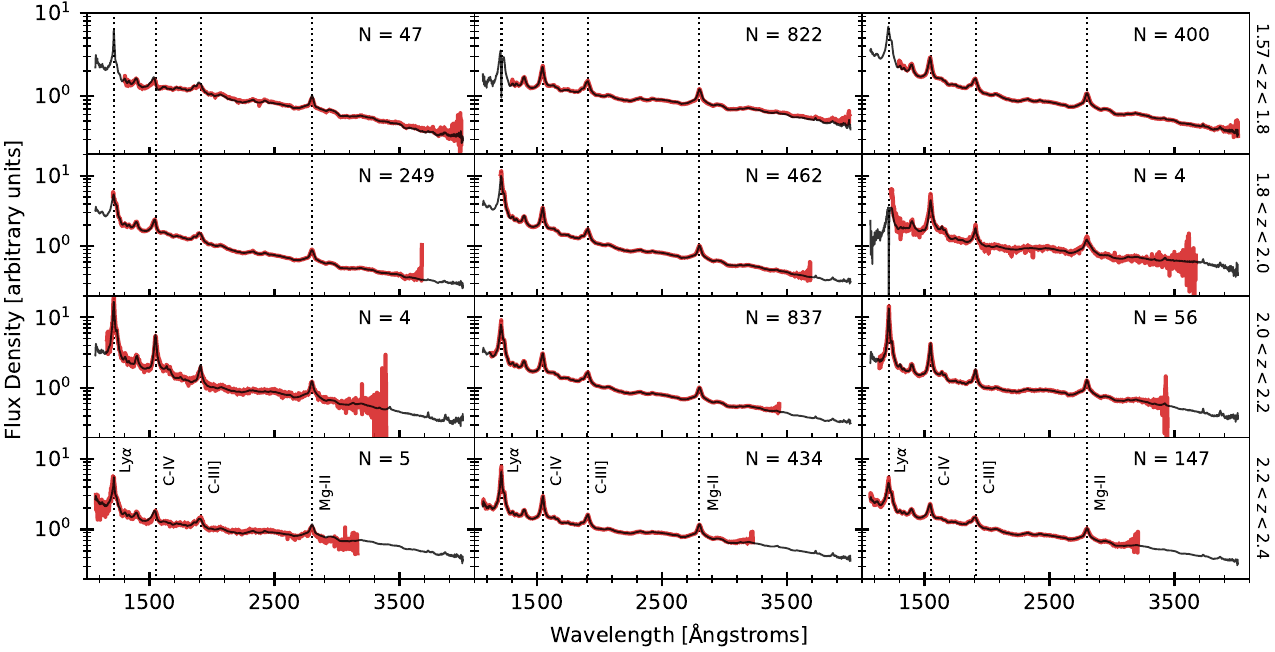}
        \caption{Three randomly
        selected composite quasar spectra from each redshift bin are shown in red. 
        The number of quasars in the cluster is noted as N.
        Overlaid in black is the fit from the 6-vector
        model described in Section~\ref{subsec:modelingTest} and its extrapolation
        (Section~\ref{subsec:PCA_extrap}) beyond the observed wavelength range.}
        \label{fig:setExamples}
\end{figure*}

One cluster composite spectrum had strong Mg~\textsc{ii} absorption features superimposed on the emission line,
resulting in a poor Gaussian fit. In this case, negative features in the continuum-subtracted flux were 
masked and the line was refit. This masking provided a substantially better estimate for the line 
location as verified by visual inspection. 

The compression of clusters results in a library containing 684 high S/N composite quasar 
spectra (see Table~\ref{table_stats}). Six quasar spectra that clustered together in the redshift bin
$1.57<z<1.8$ are shown in Figure~\ref{fig:buildEx}, along with the resulting composite spectrum. The 
spectra of this cluster all exhibit an apparent break in the continuum around the C~\textsc{iv} emission line, 
demonstrating the ability of the clustering technique to isolate trends in spectral diversity. A randomly
selected sample of composite quasar spectra from each redshift bin is shown in Figure~\ref{fig:setExamples}. 

\section{Principal Component Analysis}
We next create a spectral model to reconstruct the diversity observed in the composite
quasar spectra library. In this section, we present the eigenspectra derived from the 
library and determine the number of eigenspectra necessary to sufficiently model the 
standard test sample of spectra. We then compare the modeling performance of the 
eigenspectra to that of the B12 eigenspectra on the standard test sample, BAL test sample,
and cluster outliers.

\subsection{Eigenspectra}
\label{subsec:eigenspectra}

We use Weighted Expectation Maximization Principal Component Analysis
\citep[EMPCA\footnote{\href{https://github.com/sbailey/empca/}{https://github.com/sbailey/empca/}};][]{bailey12} 
to derive eigenspectra from the composite quasar spectra library. EMPCA was designed to
perform PCA on data sets with noise or gaps. Sources of noise in the composite quasar spectra library 
include composites from clusters with small membership and the extrema of the wavelength range
in each composite. Examples of variation in the noise levels can be seen in 
Figure~\ref{fig:setExamples}. 

The error on each co-added flux measurement was preserved when constructing the composite 
spectra. These errors are used to weight the flux measurements in each spectrum. The regions
of each spectrum with missing data over the rest-frame wavelength range $1060 - 4023$\AA\ are
given weights of zero. The error-weighted mean spectrum of the full library is subtracted from
each spectrum to maximize the variance coverage of the first eigenspectrum. Wavelength columns
below a minimum data requirement of $5 \times N$ nonzero weight values are removed, where N 
is the number of eigenspectra. We retain nine eigenspectra, covering $1067 - 4007$~\AA, and 
neglect subsequent eigenspectra.

\begin{figure*}[htb!]
  \centering
	  \includegraphics[width=0.9\textwidth, angle=0]{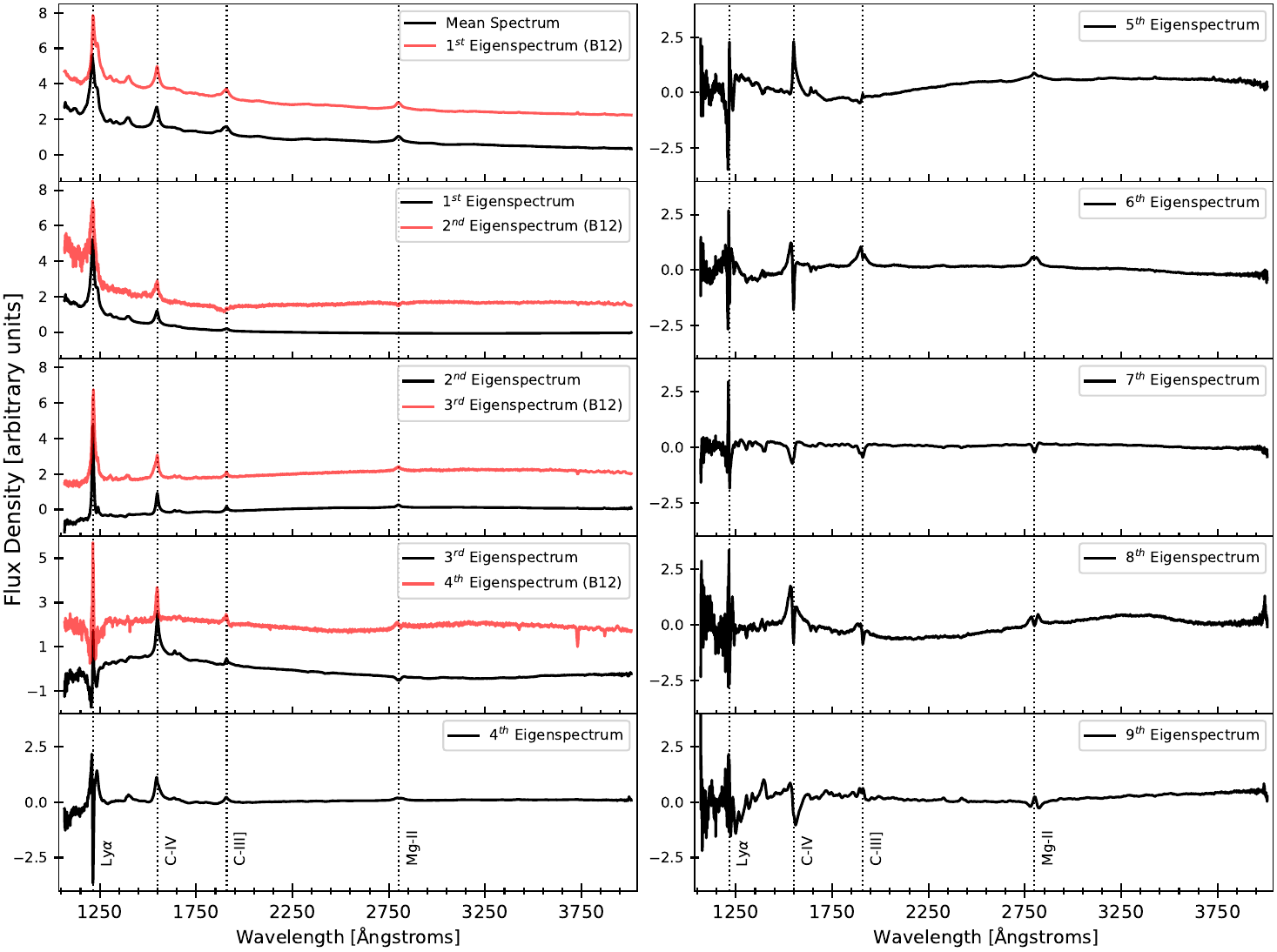}
	  \caption{The mean quasar spectrum and the first nine eigenspectra derived in 
	  this work are shown in black. The B12 eigenspectra are
	  shown in red and are offset by two flux units for illustrative purposes.}
	  \label{fig:pca}
\end{figure*}

The mean spectrum and first nine eigenspectra are shown in Figure~\ref{fig:pca}. The four B12
eigenspectra are shown in red for comparison. The spacing between pixels in both eigenspectra sets
is uniform in log wavelength at $10^{-4}$. The mean spectrum resembles the first B12 
eigenspectrum that was derived without an initial mean subtraction. There is also a noticeable 
reduction in noise for the eigenspectra of this work relative to the B12 eigenspectra. The B12 
eigenspectra are derived from 568 individual quasar spectra, leading to a lower effective
S/N.

The eigenspectra of this work capture two notable properties of the C~\textsc{iv} emission 
line: an asymmetry in the C~\textsc{iv} line profile and a shift in the line center. The peak of
C~\textsc{iv} is often significantly blueshifted from the systemic redshift of the quasar,
thought to be a result of high-velocity outflows in the broad-line region \citep[e.g.][]{richards11,shen16}.
The asymmetry is subtle in the first four eigenspectra and becomes more pronounced in the 
higher-order eigenspectra. A profile asymmetry and line center shift can also be seen in the 
C~\textsc{iii]} emission line in the fifth through eighth eigenspectrum. The ninth eigenspectrum shows 
significant noise between the Ly$\alpha$ and C~\textsc{iii]} emission lines, a region of particular 
importance for modeling high redshift quasars.

\subsection{Modeling the Standard Test Sample}
\label{subsec:modelingTest}

We use the standard test sample of 26,498 quasars to determine the number of eigenspectra
required to adequately model the data.  
Redrock\footnote{ \href{https://github.com/desihub/redrock/}{https://github.com/desihub/redrock/}} 
is used to fit the spectra with eigenspectra bases consisting of the mean and the first N 
eigenspectra, varying N from 3 to 7. Redrock is used in the same configuration to model 
these spectra using the B12 eigenspectra. Redrock searches over a specified trial redshift range 
and determines the redshift at which a linear combination of the input eigenspectra best describes
an observed spectrum. Redrock also returns the error on the redshift estimate, the coefficients of
the eigenspectra in the best fit, the $\chi^2$ of the fit, and the $\Delta\chi^2$ to the next best 
fit redshift. This last parameter is the absolute difference in $\chi^2$ between the best fit model
at the best fit redshift and the best fit model at the second best fit redshift. Larger 
$\Delta \chi^2$ values indicate a higher likelihood that the best fit 
estimate is correct. We define the trial redshift range for the test samples as $1.6167 < z < 2.3234$, 
determined by the redshift coverage of our eigenspectra. These redshift bounds permit a shift as large as $\Delta z = \pm 0.0333$ from the \verb|Z_PCA| redshifts that lie at the extrema of the test sample
redshift range.

The number of eigenspectra retained for final modeling should improve the quality of fit without providing
excessive flexibility that limits the ability to differentiate between redshift estimates. For comparison, 
\verb|Z_PCA| uses a seven-parameter model consisting of the four B12 eigenspectra plus a quadratic 
polynomial in $\log\lambda$ to absorb unmodeled broadband signal. We select
an eigenspectrum basis consisting of the mean spectrum plus the first five eigenspectra shown in 
Figure~\ref{fig:pca}, hereafter referred to as the 6-vector model. Increasing the number of eigenspectra
in the basis by 1 offers only a 1.7\% improvement on the median $\chi^2$ of the best fits. More importantly,
the inclusion of additional eigenspectra suppresses the distribution of $\Delta\chi^2$ to lower values.
The medians of the reduced $\Delta\chi^2$ distributions for a 5-, 6-, and 7-vector model are 1.440, 
1.335, and 1.120, respectively. The 16\% reduction from the 6-vector to 7-vector model is indicative 
of a poorer ability to distinguish between redshifts. 

\begin{table*}[htb]
\centering
\caption{Percentiles of the reduced $\chi^2$ distributions for the B12 and 6-vector
models on the unaltered spectra of the standard test sample and after various corrections are applied}
\begin{tabular}{c | c c c | c c c } 
 \hline
 &  & B12 & & & 6-vector & \\
 & 50th & 75th & 90th & 50th & 75th & 90th \\ 
 \hline\hline
No corrections & 1.063 & 1.254 & 1.554 & 0.972 & 1.099 & 1.301 \\ 
\hline
Polynomial terms & 1.005 & 1.155 & 1.385 & 0.945 & 1.062 & 1.255 \\ 
\hline
Extinction corrected & 1.061 & 1.252 & 1.545 & 0.972 & 1.099 & 1.300 \\
\hline
NALs filtered & 1.049 & 1.231 & 1.509 & 0.959 & 1.079 & 1.263 \\
\hline
Extinction corrected and NALs filtered & 1.047 & 1.231 & 1.501 & 0.959 & 1.079 & 1.262 \\ 

 \hline
\end{tabular}
\label{table_chi2_corrections}
\end{table*}
 
\begin{figure*}[htb!]
  \centering
	  \includegraphics[width=0.9\textwidth, angle=0]{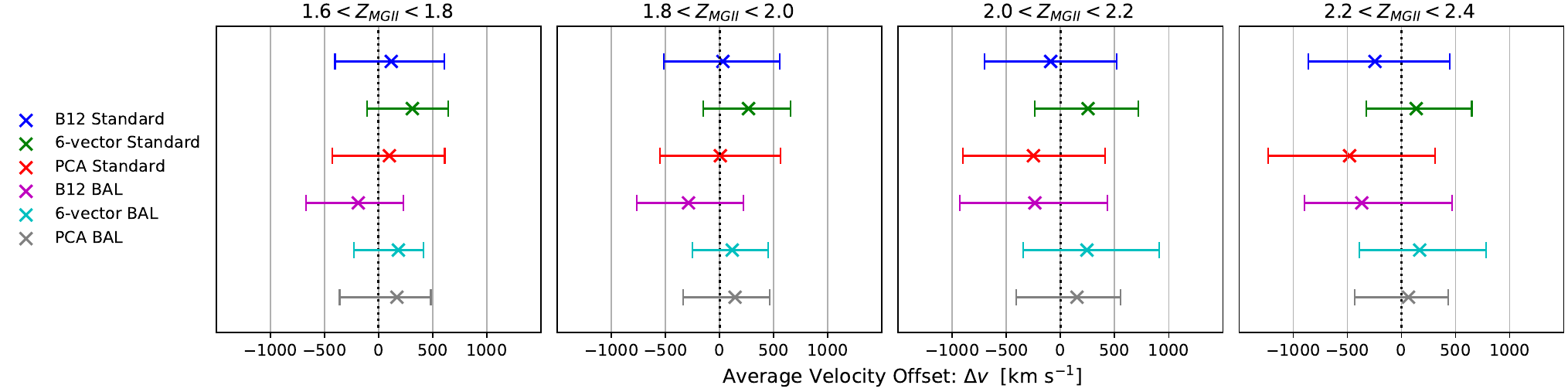}
	  \caption{The average and range of the 16th to 84th percentiles of velocity 
	  offset for each redshift bin.
	  The velocity offsets of the B12, 6-vector, and PCA redshifts
	  from Z\_MGII are shown. The difference between the
	  B12 model and PCA offsets is primarily a result of masking the Ly$\alpha$
	  forest.}
	  \label{fig:mgii_z}
\end{figure*}

We compare the performance of the 6-vector model to that of the B12 eigenspectra on the standard test
sample. We evaluate the redshift estimates, the resulting reduced $\chi^2$ distributions, and reduced 
$\Delta \chi^2$ distributions from both models. The best fits and redshift estimates derived from the
B12 eigenspectra in this work differ from \verb|Z_PCA| in a few key ways: we mask pixels blueward of
1216~\AA~in the rest frame as determined from \verb|Z_PCA| during fitting rather than correcting for
optical depth, we do not include a quadratic polynomial or apply extinction corrections to the 
spectra (we devote Section~\ref{subsec:additionalTests} to testing these corrections), and the 
redshift search window is inclusive of the range about a spectrum's \verb|Z_PCA| estimate but limited
to the wavelength coverage of the 6-vector model eigenspectra. For the complete description of 
\verb|Z_PCA|, we refer the reader to Section~4.4 of \citet{lyke20}. We reject 21 spectra that achieved
the best fit at the bounds of the trial redshift range for either the B12 or 6-vector model from all 
further analyses. These objects are primarily misclassified stars, galaxies, or quasars with a true 
redshift value outside the specified search range.

The redshift estimates from the B12 and 6-vector models are cross-checked against \verb|Z_MGII|, the 
redshift of the Mg~\textsc{ii} emission line listed in DR16Q. We also compare \verb|Z_PCA| and \verb|Z_MGII|. 
Spectra that have a warning flag associated with \verb|Z_MGII| are omitted. The velocity offset of the 
estimated redshift ($Z_{EST}$) from the B12 model, 6-vector model, and \verb|Z_PCA| relative to
\verb|Z_MGII| is determined as
\begin{equation}
\label{eq:DeltaV}
\Delta v = c \times \frac{Z_{EST} - Z_{MGII}}{1 + Z_{MGII}}
\end{equation}
The standard test sample is split on \verb|Z_MGII| into four bins of width $\Delta z = 0.2$. The
average velocity offsets in each bin are shown in Figure~\ref{fig:mgii_z} along with the corresponding
68\% confidence intervals. The 6-vector model displays a consistent, positive average velocity offset
across all redshifts. The observed offset for 6-vector redshifts might be a result of modeling choice.
We treat the Mg~\textsc{ii} emission line as Gaussian when correcting the composite spectra to the 
proper rest-frame wavelengths. The \verb|Z_MGII| redshifts are determined using the location of peak 
flux from PCA fits limited to the vicinity of the Mg~\textsc{ii} line. 

The average velocity offset across all redshift bins ranges from 136 to 312 km s$^{-1}$ for the 6-vector 
model and from $-$245 to 117 km s$^{-1}$  for the B12 model. The average redshift estimates from the B12 model
are more consistent with \verb|Z_MGII| than the 6-vector model estimates in the three lower redshift bins
but, similar to \verb|Z_PCA|, trend lower with redshift and exhibit greater scatter about the average at 
all redshifts. The average confidence interval has a width of 869 km s$^{-1}$ for the 6-vector model and
1144 km s$^{-1}$ for the B12 model.

The coefficients of the 6-vector model eigenspectra display one notable trend with absolute velocity 
offset from \verb|Z_MGII|. The coefficients of the second eigenspectrum become more negative, while 
the coefficients of the fifth eigenspectrum trend positive with increasing offset. This trend may be
related to the blueshifting of high-ionization emission lines. The greater flux density in these
lines relative to the more stable Mg~\textsc{ii} line tends to control PCA fits when present.
As these lines shift from the systemic redshift, the second eigenspectrum, dominated by narrow line
cores, reduces flux at the line centers. Meanwhile, the fifth eigenspectrum, clearly linked to red-blue 
asymmetry in the C~\textsc{iv} and C~\textsc{iii} line profiles, acts to correct the shape of the
lines, possibly in combination with other eigenspectra.

There are 14 instances where the redshift estimates from the B12 and 6-vector models differ by 
$\Delta v > 10,000$~km~s$^{-1}$. We visually inspect the spectra to identify the source of tension.
Eight of these instances are misclassified stellar spectra or quasar spectra with unidentified BALs.
The 6-vector model is consistent with the visual inspection redshift in two cases, and the B12 model 
is consistent with the visual inspection redshift in three cases. There is one instance were both 
models failed to accurately estimate the redshift. In addition to the B12 and 6-vector model redshift 
disagreements, there are 13 spectra with redshift disagreements between the 6-vector model and 
\verb|Z_PCA| of $\Delta v > 10,000$~km~s$^{-1}$. These spectra include the three cases where the B12 model
correctly estimates the redshift and seven of the misclassified spectra discussed previously. For the 
remaining three spectra, one is a misclassified nonquasar spectrum, one has an incorrect redshift 
from both the 6-vector model and \verb|Z_PCA|, and one has a correct \verb|Z_PCA| redshift estimate but
incorrect 6-vector model redshift.

\begin{figure}[htb!]
  \centering
	  \includegraphics[width=0.45\textwidth, angle=0]{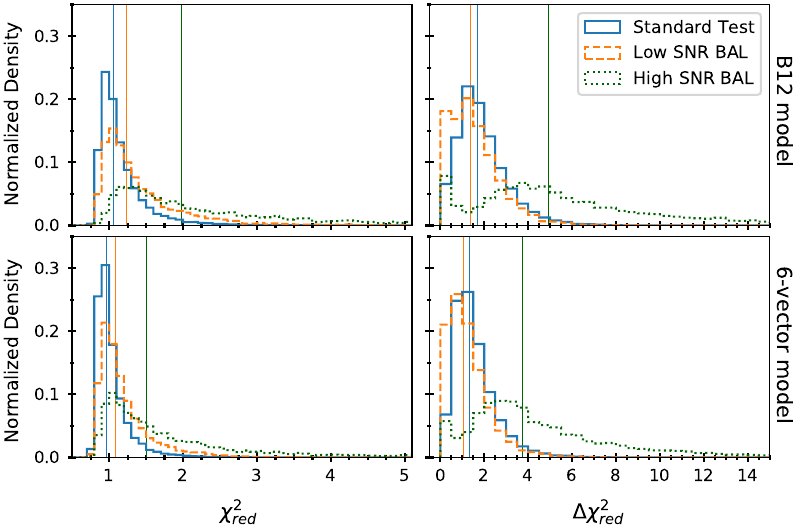}
	  \includegraphics[width=0.43\textwidth, angle=0]{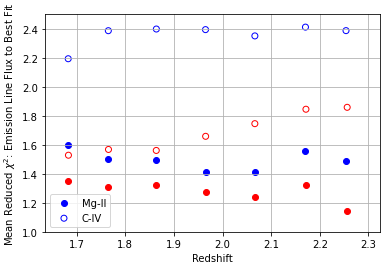}
	  \caption{\textbf{Top:} the distributions of reduced $\chi^2$ for the best fits
	  to the spectra of the standard test sample and BAL test sample using the B12 (upper left) and
	  6-vector (lower left) models, and the corresponding distributions of
	  reduced $\Delta\chi^2$ for the B12 (upper right) and 6-vector (lower right) models. 
	  In all panels, the BAL test sample is split into two
	  subsamples at an average S/N per pixel of 10. The vertical
	  lines are the median value of each distribution.
	  \textbf{Bottom:} the reduced $\chi^2$ of the best fits from the B12 (blue)
	  and 6-vector (red) models in the region of the Mg~\textsc{ii} 
	  emission line and C~\textsc{iv} emission line in the standard test
	  sample as a function of redshift. The uncertainties on the average reduced $\chi^2$
	  are on the 1\% level and omitted for clarity.}
	  \label{fig:chi2distribution}
\end{figure}

The distributions of reduced $\chi^2$ and reduced $\Delta\chi^2$ from the B12 and 6-vector models are
shown in Figure~\ref{fig:chi2distribution}. The 50th, 75th, and 90th percentiles of the reduced $\chi^2$
distributions from each model are given in the first row of Table~\ref{table_chi2_corrections}. The more
compact reduced $\chi^2$ and reduced $\Delta \chi^2$ distributions from the 6-vector model relative to the B12
model imply that the 6-vector model provides a better fit to the observed data, but the B12 model is more 
confident in the redshift estimate, on average. A reduced $\Delta\chi^2$ of less than 0.01 is considered the threshold
for a poorly determined redshift. Applying this threshold to the reduced $\Delta\chi^2$ distributions, we 
find that 0.72\% of redshifts from the B12 model are poorly determined, compared to 0.45\% from the 6-vector
model.

The composite quasar spectra library was built using the assumption that the location of
Mg~\textsc{ii} emission line is the least biased from the systemic redshift. We therefore
investigate the modeling performance specifically in the Mg~\textsc{ii} emission line region.
Using the best fits to the full spectrum for the B12 and 6-vector models, we determine the 
reduced $\chi^2$ over the range $2750 - 2850$~\AA\ for spectra with 50 or more good pixels in
this range. The results are shown in Figure~\ref{fig:chi2distribution}. The 6-vector model 
provides lower $\chi^2$ fits to the Mg~\textsc{ii} region, at 9\% on average, across the entire
redshift range. 

We repeat the above procedure for the C~\textsc{iv} emission line region over the
wavelength range of $1500 - 1600$~\AA. The results are also shown in Figure~\ref{fig:chi2distribution}.
The difference in reduced $\chi^2$ between the models is more pronounced in the C~\textsc{iv} 
region. Visual inspection reveals that the B12 model is frequently underestimating the peak flux 
measurements of the C~\textsc{iv} emission line. We prioritize the information contained 
in emission line regions in our clustering analysis, placing emphasis on capturing the 
range of diversity. Additionally, we are able to expand the basis to six spectral templates 
instead of four owing to the high S/N of our sample. Both of these factors 
facilitate reconstruction of the emission line diversity.
The trend of increasing reduced $\chi^2$ in the C~\textsc{iv} region with redshift for
the 6-vector model is likely due to the presence of the Ly$\alpha$ emission line influencing 
the overall fit to quasar spectra with $z>2$.

\subsubsection{Additional Modeling Tests}
\label{subsec:additionalTests}

In the previous subsection, we simply modeled the processed spectra with linear combinations
of the eigenspectra, excluding flux measurements in the Ly$\alpha$ forest. We now explore the
various corrections that were applied to the training sample to assess potential benefits to 
spectral modeling with the B12 and 6-vector models. We evaluate how correcting for galactic 
extinction, filtering NALs, and including a quadratic polynomial in $\lambda$ in the fit to 
absorb broadband signal affect the quality of the best fits on the standard test sample.

\begin{figure*}[htb]
    \centering
        \includegraphics[width=0.85\textwidth, angle=0]{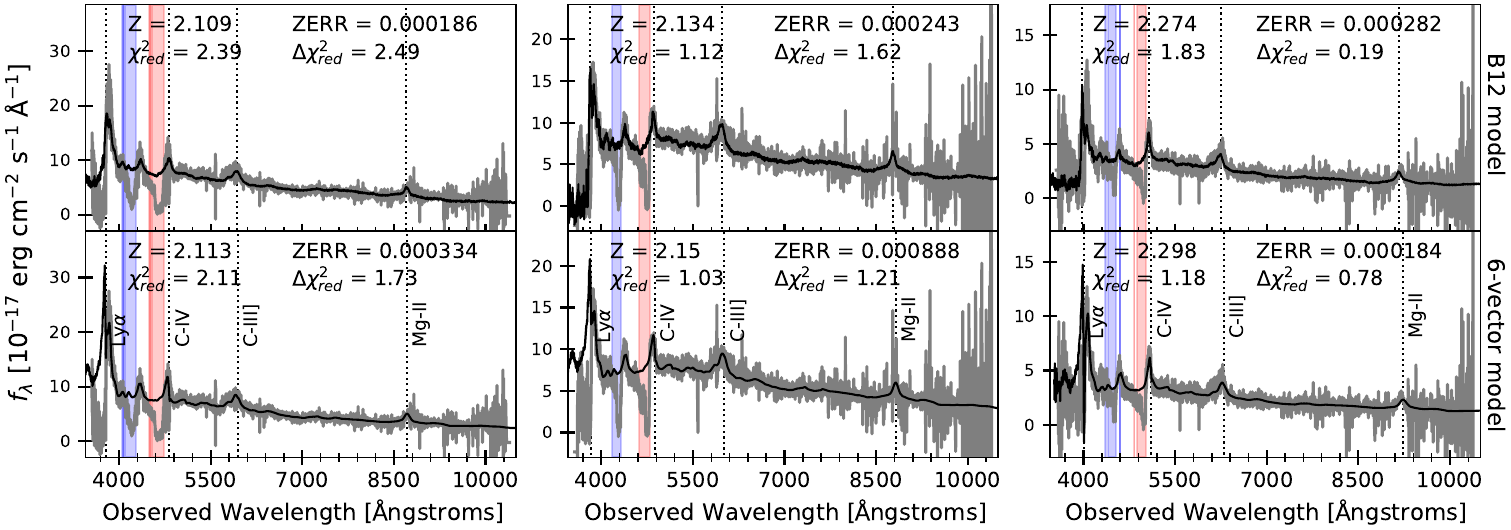}
        \caption{Each column shows the observer-frame spectrum of a
        randomly selected BAL quasar in gray. The fits from the B12 (top) and 6-vector (bottom) models
	    are overlaid on the observed spectrum in black. 
        The red and blue shaded regions indicate the masked C~\textsc{iv} and Si~\textsc{iv} BAL 
        troughs, respectively, as recorded in the BAL Quasar Catalog. 
        The spectra have a BAL\_PROB of at least 0.9.}
\label{fig:BAL_examplePCAFits}
\end{figure*}

Table~\ref{table_chi2_corrections} quantifies the reduced $\chi^2$ distributions of the 
spectral fits from the B12 and 6-vector models when the corrections to the spectra are applied. 
The row for ``No Corrections'' corresponds to the initial best fits. Both models are minimally 
effected by extinction correction and filtering NALs. The NAL filter simply removes outlier pixels
from each spectrum, so both models experience comparable gains. The B12 model benefits significantly from 
the addition of the quadratic polynomial, while the 6-vector model achieves at most a 4\% improvement
for any of the corrections. In all iterations, the 6-vector model achieves smaller reduced
$\chi^2$ values than the B12 model at each percentile. All further modeling is performed without 
the additional corrections or polynomial tested in this section.

\subsection{Modeling the BAL Test Sample}
\label{subsec:modelBAL}

We next evaluate the ability of the 6-vector model and B12 model to fit the quasar spectra
in the BAL test sample. We reject the spectra for which Redrock achieves the best fit at the
redshift bounds for at least one of the models from all further analyses.

Figure~\ref{fig:BAL_examplePCAFits} shows three randomly selected BAL quasar spectra and the
best fits from the B12 and 6-vector models. The selected quasars all have a \verb|BAL_PROB| 
of greater than 0.9. The regions identified in the BAL Quasar Catalog as contaminated with 
C~\textsc{iv} or Si~\textsc{iv} associated BALs are indicated in the figure. Both models 
adequately fit the unabsorbed regions of each spectrum. 

We split the BAL test sample on average S/N per pixel at a value of 10, motivated by 
the S/N limit of the standard test sample. The distribution of the reduced $\chi^2$ and 
reduced $\Delta\chi^2$ for the B12 and 6-vector models on the low and high S/N BAL test
samples is shown in Figure~\ref{fig:chi2distribution}. 

The similarity of the reduced $\chi^2$ distributions on the low S/N 
BAL sample and the standard test sample validates our approach to masking the absorption 
features and supports that BAL and non-BAL quasars are not distinct objects
\citep[consistent with][]{rankine20}. The reduced $\chi^2$ distribution for both models on the high 
S/N BAL test sample, however, trends toward higher values, with medians of 1.980 and
1.507 for the B12 and 6-vector model, respectively. BALs can be associated with several different
emission lines, yet we only mask the most commonly observed BALs associated with C~\textsc{iv}
and Si~\textsc{iv}. The high S/N spectra will suffer a larger penalty than low 
S/N spectra in the $\chi^2$ statistic by not properly modeling other BAL features, if 
present. In addition, it is impossible to perfectly identify and mask all the C~\textsc{iv} and 
Si~\textsc{iv} BAL features that occur; thus, some will unavoidably slip by the masks.

The distribution of $\Delta\chi^2$ for the low S/N BAL test sample resembles that for
the standard test sample for both the B12 and 6-vector models. As with the standard test sample, 
the B12 model typically achieves larger $\Delta\chi^2$ values for both BAL test samples relative 
to the 6-vector model. The high S/N sample has a bimodal distribution with a first peak near
zero for both models. Applying the threshold of reduced $\Delta \chi^2 < 0.01$ for poorly determined
redshift, 0.79\% and 0.73\% of redshifts are poorly determined from the 6-vector and B12 models, 
respectively, for the entire BAL sample.

\begin{figure*}[htb!]
  \centering
	  \includegraphics[width=0.85\textwidth, angle=0]{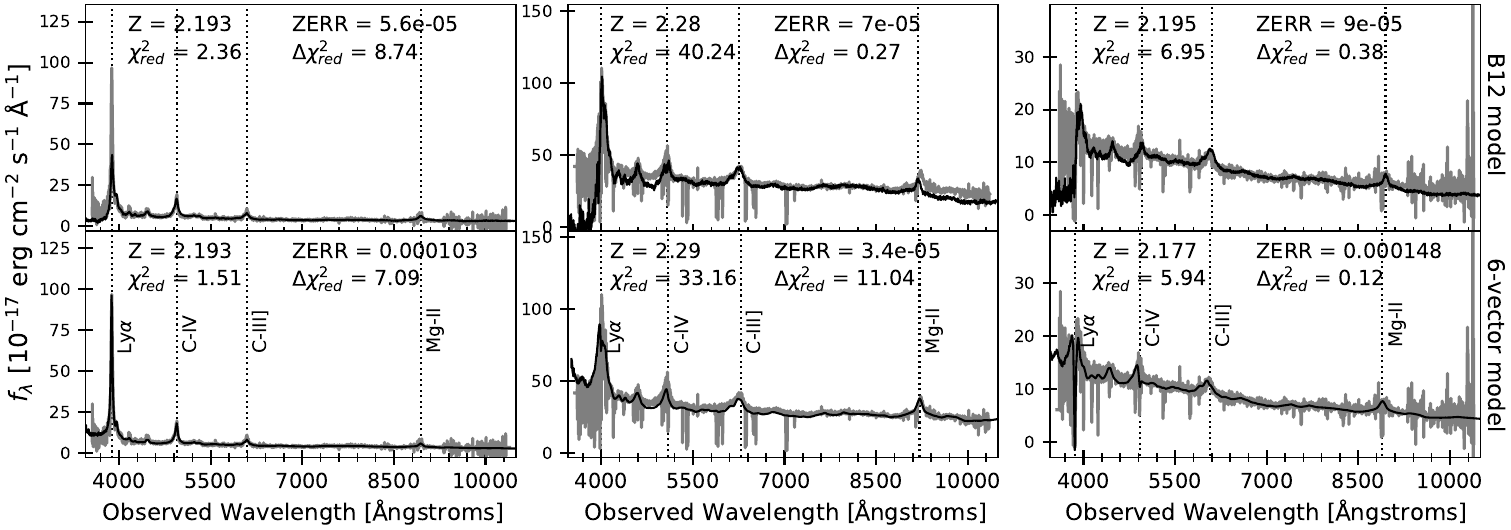}
	  \caption{Each column shows the observer-frame spectrum of a randomly 
	  selected cluster outlier in gray.
	  The fits from the B12 (top) and 6-vector (bottom) models
	  are overlaid on the observed spectrum in black.}
	  \label{fig:outliers_PCAfit}
\end{figure*}

Following the same procedure as with the standard test sample, the redshift estimates from the 
B12 model, the 6-vector model, and \verb|Z_PCA| are checked against \verb|Z_MGII|. The results 
are shown in Figure~\ref{fig:mgii_z}. As with the standard test sample, the average offsets from 
the 6-vector model are consistently positive, show less redshift dependence, and have smaller 
confidence intervals compared to the B12 model across all bins. The average velocity offset ranges 
from 117 to 244 km s$^{-1}$ for the 6-vector model and from -366 to -188 km s$^{-1}$  for the B12
model, with average confidence interval widths of 916 and 1158 km s$^{-1}$, respectively. 
The B12 and 6-vector models show more scatter in the redshift estimates for BAL quasars 
at high redshifts compared to non-BAL quasars as indicated by the size of the confidence
intervals for the two highest redshift bins.

There are 60 spectra with redshift estimates from the B12 and 6-vector models that differ
by more than 10,000 km s$^{-1}$. We visually inspect these spectra and find that the 6-vector model
correctly estimates the redshift in 18 cases, the B12 model correctly estimates the redshift in
11 cases, and both models fail to estimate the redshift in 13 cases. For the spectra where the 
B12 model redshift is correct, the 6-vector model most often inverts one or more emission lines
to match an unmasked BAL feature. For the spectra where the 6-vector model redshift is correct,
the B12 model most often exhibits line confusion, matching the template C~\textsc{iv} and 
C~\textsc{iii]} emission lines to the true Ly$\alpha$ and C~\textsc{iv} flux. Additionally, there
are 10 instances of misclassified galaxies, stars, and low redshift quasars. The remaining eight 
spectra are FeLoBAL quasars, a rare class of spectra exhibiting BALs from low-ionization species
including Fe~\textsc{ii} and Fe~\textsc{iii} \citep{hall02}.

\subsection{Modeling the Cluster Outliers}
\label{subsec:modelOutliers}

We assess the ability of the B12 and 6-vector models to fit more peculiar spectra by fitting a 
selection of the cluster outlier spectra described in Section~\ref{subsec:outliers}. Of the 403 outliers,
271 cluster outliers confirmed as quasars within the redshift range of the standard test sample 
($1.65<z<2.29$) are modeled. One quasar spectrum in this sample has an underestimated \verb|Z_PCA| 
redshift with a true value beyond the search window and is excluded.

Three cluster outlier spectra and the best fits from the B12 and 6-vector models are shown in 
Figure~\ref{fig:outliers_PCAfit}. While the reduced $\chi^2$ values are large compared to the 
standard test sample, both models provide a qualitatively good fit to the data and are consistent
in redshift estimate.

The distributions of reduced $\chi^2$ are generally degraded relative to the standard test
sample, with median values of 4.871 and 3.480 for the B12 and 6-vector model, respectively.
The majority of the modeled outlier spectra contain excessive, unmasked NALs or strong 
intrinsic absorption features. In nearly all these cases, the best fit from both models 
describes the observed data well and provides accurate redshift estimates verified by visual 
inspection. The higher reduced $\chi^2$ distributions relative to the standard test sample likely 
result from the same absorption features that caused the spectra to be rejected as outliers. We 
note only one redshift failure for these types of outlier spectra in which the 6-vector model
inverts an emission feature to match observed absorption in the spectrum. For spectra with strong 
intrinsic absorption or NALs superimposed on the Ly$\alpha$ emission line, there is a notable 
trend of the fourth and fifth eigenspectra of the 6-vector model becoming more dominant.
The increased contribution of these eigenspectra in the best fit is evident as a sharp decrement
in flux at the Ly$\alpha$ line center and asymmetry in the C~\textsc{iv} line profile of the fit
that is not apparent in the observed spectrum (e.g. third column of Figure~\ref{fig:outliers_PCAfit}). 
The deep flux decrement at the Ly$\alpha$ line center in the fourth and fifth eigenspectra results from
strong intrinsic absorption features in a few of the composite spectra from which the principal
components were derived. For future studies, a mask may need to be applied to the composite spectra
at the location of these strong absorption features to prevent nonphysicality in modeling. 

In Section~\ref{subsec:outliers}, we identified a broad category of outliers with peculiar spectral
features. Most outlier spectra identified as peculiar have atypical emission line flux ratios, 
continuum slopes, or breaks in spectral index. Both models unambiguously provide accurate redshifts
and qualitatively good fits for these spectra. The B12 model often underestimates the emission line
flux for Ly$\alpha$, Si~\textsc{iv}, C~\textsc{iv}, and C~\textsc{iii]} across the full outlier sample
and in spectra with extreme line ratios (see, e.g., first column of 
Figure~\ref{fig:outliers_PCAfit}). In spectra with a particularly strong and narrow Ly$\alpha$ emission 
line, the 6-vector model also often falls short in estimating the peak flux. As for spectra with a break
in spectral index, the 6-vector model captures the continuum flux more accurately in the far red than 
does the B12 model.

In addition to the peculiar spectra discussed above, we find a small subset of spectra with 
unidentified C~\textsc{iv} and Si~\textsc{iv} BALs and one FeLoBAL spectrum. The results 
for spectra with unidentified C~\textsc{iv} and Si~\textsc{iv} BALs are mixed for both models. We 
do not discuss these further, as the previous section provides a detailed discussion on modeling the 
spectra of high ionization BAL quasars. For the FeLoBAL spectrum, both models provide accurate 
redshift estimates, but the continuum fits are influenced heavily by the presence of absorption.

Lastly, there are three peculiar outlier quasars with extremely reddened spectra where the redshift
is difficult to estimate. Both models fit one spectrum with clear C~\textsc{iv}, C~\textsc{iii]}, and
Mg~\textsc{ii} emission lines surprisingly well with accurate redshift estimates. We do not evaluate the 
best fits to the other two dust-reddened spectra owing to uncertainty in the true redshift. In future
implementation of this modeling method, we will consider supplementing the templates to better 
inform the fitting of dust-reddened quasars.

\subsection{PCA Extrapolation and Archetypes}
\label{subsec:PCA_extrap}

Through fitting an observed spectrum with a linear combination of eigenspectra, we can 
extrapolate beyond the observed wavelength bounds to the full wavelength range of the PCA model.
We fit the spectra of the composite library with the 6-vector model and use the best fits to 
predict each spectrum's behavior over the wavelength coverage of the eigenspectra. We show the
best fits and extrapolation from the 6-vector model to a random selection of composite spectra in
Figure~\ref{fig:setExamples}.

The models derived from the composite quasar spectra library adequately capture the range of
intrinsic diversity in the test samples; therefore, we suggest that the set of extended PCA model 
spectra from the composite quasar spectra library are accurate representations of quasar
spectral diversity. The extended models can be interpreted as essentially noiseless, 
data-driven archetypes.

There are many potential applications for a quasar archetype library. In particular, if the 
redshift range for this study is expanded, the archetypes could inform studies of the Ly$\alpha$
forest. It may be feasible to predict the unabsorbed flux in regions contaminated by the forest
through matching an observed spectrum to an archetype. Alternatively, measuring the fraction of 
quasars represented by a given archetype as a function of redshift can provide information on the
redshift evolution of quasars.

\section{Discussion}
We performed a clustering analysis on 12,968 quasar spectra in the redshift range of $1.57<z<2.4$, 
compressing the sample by nearly a factor of 12. We first determined archetype spectra that are
representative of the entire quasar sample. We then identified clusters of quasar spectra, each centered
on an archetype spectrum, that are similar on the level of individual flux densities. We implemented
three modifications specific to the peculiarities of a large quasar sample. We removed foreground 
absorption features from the spectra. We reduced the effect of broadband signal in the distance
measurement assuming a power law. Lastly, we focused comparisons on the spectral regions
that display the most variation, primarily prominent emission lines. We label individual spectra 
that could not be assigned to any cluster as outliers. We then constructed a library of high 
S/N composite quasar spectra from the clusters. Including the outliers, our final,
compressed sample contains 684 composite quasar spectra, 5 clusters of misclassified stars,
1 misclassified galaxy spectrum, 6 quasar spectra with incorrect redshift estimates resulting 
from line confusion in PCA, and 396 quasar spectra that are unlike the rest of the sample for 
varying reasons. 

We tested whether the library is representative of the spectral diversity over the sampled redshift range by
developing a spectral model through PCA. Our model, the 6-vector model, consists of the mean spectrum 
of the composite quasar spectra library and the first five eigenspectra. We fit three independent test 
samples with the 6-vector model: a standard sample of non-BAL quasar spectra, a BAL quasar spectra sample with
identified C~\textsc{iv} and Si~\textsc{iv} BAL features masked, and a selection of outlier spectra from
the clustering analysis. All samples were restricted to the redshift range of $1.65<z<2.29$ to permit a 
free redshift parameter in modeling. The best fits from the 6-vector model were compared to the best fits
from the B12 eigenspectra. The results are summarized as follows:

\begin{itemize}
    \item The 6-vector model achieves an 8.5\% reduction in median $\chi^2$ on the standard test sample relative to the B12 model. The 6-vector model shows significant improvement in the tail of the reduced $\chi^2$ distribution for all samples (see Table~\ref{table_chi2_corrections}).
    \item On average, the B12 model achieves higher values of $\Delta\chi^2$ than the 6-vector model, indicative of stronger discrimination between competing redshift estimates. We find that 0.72\% of B12 redshift estimates and 0.45\% of 6-vector redshift estimates fail to meet the threshold of $\Delta\chi^2>0.01$ for the standard test sample. The percentages of poorly determined redshifts on the BAL test sample are roughly equal for both models.
    \item Both the B12 and 6-vector models attain best fit statistics on the BAL test sample that are comparable to the standard test sample over the same range of S/N. These results support the theory that BAL quasars are not distinct objects and that they display the same range of intrinsic variation in the unabsorbed regions of their spectra as non-BAL quasars.
    \item Both the B12 and 6-vector models achieve redshift estimates consistent with the redshift of the Mg~\textsc{ii} emission line. The B12 model shows stronger average consistency with the Mg~\textsc{ii} redshift at lower redshifts but less self-consistency and greater redshift-dependent bias than the 6-vector model across all redshifts. 
    \item The 6-vector model provides higher quality fits to the regions of the C~\textsc{iv} and Mg~\textsc{ii} emission lines than the B12 model, as shown in Figure~\ref{fig:chi2distribution}. The B12 model often underestimates C~\textsc{iv} emission line flux, revealed by visual inspection.
    \item Both models provide good redshift estimates and qualitatively good descriptions of the spectra in the outlier sample. The 6-vector model appears more equipped to model the flux in spectra with extreme emission line ratios. However, there is an unphysical trend in the 6-vector model that leads to occasionally sharp flux decrements on the Ly$\alpha$ emission line. The sharp feature appears in the fourth and fifth eigenspectra as shown in Figure~\ref{fig:pca}.
\end{itemize}

Several factors contribute to the enhanced performance of the 6-vector model relative to B12.
The PCA training sample is constructed to capture the range of population characteristics, 
primarily in emission line regions, while rejecting contaminants such as misclassified objects.
This focus better equips the model to handle atypical emission line fluxes and ratios, as noted in 
the last bullet point. Further, the high S/N of the sample allowed for correcting all 
composite spectra to the proper rest frame of the Mg~\textsc{ii} emission line through direct fitting
of the line. This correction leads to the significant reduction in redshift-dependent bias when
redshifting spectra with the 6-vector model. Of equal importance to the previous factors listed, 
we use 20 times more quasars than the B12 sample, producing eigenspectra with significantly less 
noise. The noise levels permit retention of higher-order eigenspectra for fitting.

We now summarize some improvements that could further optimize the performance of this technique. 
First, a more rigorous NAL filter should reduce the number of quasar spectra that are rejected as outliers.
A secondary absorption line filter could be applied to the composite quasar spectra library as well, 
potentially reducing the nonphysical trends we observe in the higher-order eigenspectra. Lastly, 
we used the propagated error on the composite spectra as weights in EMPCA. We may adjust this approach
to increase representation of the smaller clusters and improve modeling of more rare objects.

In the next stages of this study, we will expand the redshift range using the data from the first few months
of DESI observations. We echo the sentiments of \citet{yip04} that multiple eigenspectra sets, each spanning 
separate redshift ranges, will improve modeling. Eigenspectra sets covering a specific redshift range can be 
tuned to the relevant features, consequently providing a proper description of the spectral diversity and
allowing for redshift evolution. A natural choice for a split in eigenspectra set is $z\approx1.5$, below which
BALs cannot be automatically detected owing to the absence of the C~\textsc{iv} line. BAL diversity could be folded 
into clustering at low redshifts, allowing spectral templates the flexibility to model these features. Systemic
redshifts can be trained on [O~\textsc{ii}] and Mg~\textsc{ii} for the low and high redshift templates, respectively.
Deriving independent spectral models for these two redshift ranges will also allow us to properly capture the
spectral diversity arising from galaxy features in the low redshift sample and from the most luminous quasars
in the high redshift sample. This new series of spectral models could reduce systematic errors in redshift 
estimates, mitigate the impact of BALs, and improve predictions of the Ly$\alpha$ forest 
continuum, thus leading to improved measurements of  the baryon acoustic oscillation feature and redshift space
distortion with the DESI quasar sample.

\section*{ACKNOWLEDGEMENTS}
{The authors wish to thank Stephen Bailey for very helpful discussions. The work of Allyson Brodzeller and Kyle Dawson was supported in part by U.S. Department of Energy, Office of Science, Office of High Energy Physics, under Award No. DESC0009959.

Funding for the Sloan Digital Sky 
Survey IV has been provided by the 
Alfred P. Sloan Foundation, the U.S. 
Department of Energy Office of 
Science, and the Participating 
Institutions. 

SDSS-IV acknowledges support and 
resources from the Center for High 
Performance Computing  at the 
University of Utah. The SDSS 
website is www.sdss.org.

SDSS-IV is managed by the 
Astrophysical Research Consortium 
for the Participating Institutions 
of the SDSS Collaboration including 
the Brazilian Participation Group, 
the Carnegie Institution for Science, 
Carnegie Mellon University, Center for 
Astrophysics | Harvard \& 
Smithsonian, the Chilean Participation 
Group, the French Participation Group, 
Instituto de Astrof\'isica de 
Canarias, The Johns Hopkins 
University, Kavli Institute for the 
Physics and Mathematics of the 
Universe (IPMU) / University of 
Tokyo, the Korean Participation Group, 
Lawrence Berkeley National Laboratory, 
Leibniz Institut f\"ur Astrophysik 
Potsdam (AIP),  Max-Planck-Institut 
f\"ur Astronomie (MPIA Heidelberg), 
Max-Planck-Institut f\"ur 
Astrophysik (MPA Garching), 
Max-Planck-Institut f\"ur 
Extraterrestrische Physik (MPE), 
National Astronomical Observatories of 
China, New Mexico State University, 
New York University, University of 
Notre Dame, Observat\'ario 
Nacional / MCTI, The Ohio State 
University, Pennsylvania State 
University, Shanghai 
Astronomical Observatory, United 
Kingdom Participation Group, 
Universidad Nacional Aut\'onoma 
de M\'exico, University of Arizona, 
University of Colorado Boulder, 
University of Oxford, University of 
Portsmouth, University of Utah, 
University of Virginia, University 
of Washington, University of 
Wisconsin, Vanderbilt University, 
and Yale University.
}

\bibliographystyle{aasjournal}
\bibliography{references}

\end{document}